\documentclass[pamq,keywordsasfootnote,reqno]{ipart}

\RequirePackage{hyperref}

\startlocaldefs
\endlocaldefs

\pubyear{2019}
\volume{0}
\issue{0}
\firstpage{1}
\lastpage{1}

\usepackage{cite}
\usepackage{bbm}
\usepackage{url}
\usepackage{color}
\usepackage{graphicx}
\usepackage{amssymb,latexsym}
\usepackage{tensor}
\usepackage[normalem]{ulem}
\usepackage{slantsc}

\usepackage{mathtools}

\usepackage{marginnote}
\usepackage{mathrsfs}
\usepackage{enumitem}

\newcommand{\proof}{\noindent {\sc Proof:}}
\newcommand{\qed}{\hfill $\Box$}

\newcommand{\pamq}[1]{}
\renewcommand{\pamq}[1]{#1}
\renewcommand{\pamq}[1]{\ptc{the red text only in the pamq version} \red{#1}}

\newcommand{\absV}{V}

\newcommand{\wmcQ}{\quadratic{\mathcal{Q}}}

\newcommand{\fancyD}{{\mycal{D}}}
\newcommand{\fancyL}{{\mycal{L}}}

\definecolor{applegreen}{rgb}{0.55, 0.71, 0.0}
\definecolor{armygreen}{rgb}{0.29, 0.33, 0.13}

\definecolor{caribbeangreen}{rgb}{0.0, 0.8, 0.6}

\newtheorem{thm}{Theorem}[section]
\newtheorem{theorem}[thm]{Theorem}

\newtheorem{remark}[thm]{Remark}

\newcommand{\bhatpsi}{ {\hat{\psi}}}
\newcommand{\bcheckpsi}{ {\check{\psi}}}

\newcommand{\sqrtbg}{{d\mu_{\backg}}}%
\newcommand{\rsqrtbg}{}%

\newcommand{\rrho}{{\sigma}}

\newcommand{\care}[2]{\underbrace{#2}_{\mbox{\color{blue}\rm \scriptsize #1}}}

\definecolor{orange}{RGB}{255,127,0}

\newcommand{\quadratic}[1]{ {#1}}

\newcommand{\gauge}[1]{ {#1}}

\newcommand{\divergence}[1]{ {#1}}

\newcommand{\ptcheck}[1]{\ptc{checked on #1}}
\renewcommand{\ptcheck}[1]{}

\newcommand{\sqrdof}[1]{|#1|^2_{\backg}}

\usepackage{tikz}

\DeclareFontFamily{OT1}{rsfs}{}
\DeclareFontShape{OT1}{rsfs}{m}{n}{ <-7> rsfs5 <7-10> rsfs7 <10-> rsfs10}{}
\DeclareMathAlphabet{\mycal}{OT1}{rsfs}{m}{n}

{\catcode `\@=11 \global\let\AddToReset=\@addtoreset}
\AddToReset{equation}{section}

\newcounter{mnotecount}[section]

\newcommand{\red}[1]{{\color{red}#1}}

\newcommand{\hdherr}{O\left( |h|^2_{\backg}\, |\bcov h|_{\backg}\right)}
\newcommand{\hdhsqerr}{O\left( |h|_{\backg}\, |\bcov h|^2_{\backg}\right)}
\newcommand{\herr}{O\left( |h|^3_{\backg}\right)}

\newcommand{\adsR}[1]{\overline{R}\tensor{\vphantom{R}}{#1}}%
\newcommand{\fullR}[1]{R\tensor{\vphantom{R}}{#1}}
 \newcommand{\diffGamma}[1]{\tensor{\delta \Gamma}{#1}}

\newcommand{\back}[1]{\overline{#1}} %
\newcommand{\backg}{\overline{g}} %
\newcommand{\bcov}{\back{D}} %

\newcommand{\bR}{\back{R}} %

\newcommand{\eel}[1]{\label{#1}\end{equation}}
\newcommand{\eeal}[1]{\label{#1}\end{eqnarray}}

\newcommand{\bel}[1]{\begin{equation}\label{#1}}
\newcommand{\bea}{\begin{eqnarray}}
\newcommand{\bean}{\begin{eqnarray}\nonumber}
\newcommand{\beal}[1]{\begin{eqnarray}\label{#1}}
\newcommand{\eea}{\end{eqnarray}}

\newcommand{\nn}{\nonumber}

\newcommand{\be}{\begin{equation}}
\newcommand{\eeq}{\end{equation}}
\newcommand{\ee}{\end{equation}}
\newcommand{\beqa}{\begin{eqnarray}}
\newcommand{\eeqa}{\end{eqnarray}}
\newcommand{\beqan}{\begin{eqnarray*}}
\newcommand{\eeqan}{\end{eqnarray*}}
\newcommand{\ba}{\begin{array}}
\newcommand{\ea}{\end{array}}

\newcommand{\mnote}[1]
{\protect{\stepcounter{mnotecount}}$^{\mbox{\footnotesize
$
\bullet$\themnotecount}}$ \marginpar{%
\raggedright\tiny\em
$\!\!\!\!\!\!\,\bullet$\themnotecount: #1} }

\newcommand{\warn}[1]
{\protect{\stepcounter{mnotecount}}$^{\mbox{\footnotesize
$
\bullet$\themnotecount}}$ \marginpar{%
\raggedright\tiny\em
$\!\!\!\!\!\!\,\bullet$\themnotecount: {\bf Warning:} #1} }

\newcommand{\R}{\mathbb R}

\newcommand{\eq}[1]{(\ref{#1})}

\newcommand{\ptc}[1]{\mnote{{\bf }#1}}

\newcommand{\beaa}{\begin{eqnarray*}}
\newcommand{\eeaa}{\end{eqnarray*}}

\def\ben{\begin{equation}}
\def\een{\end{equation}}
\def\bena{\begin{eqnarray}}
\def\eena{\end{eqnarray}}

\def\f(#1/#2){\frac{#1}{#2}}
\def\Frac(#1/#2){\left(\frac{#1}{#2}\right)}
\def\chris(#1-#2-#3){{\mit \Gamma}^{#1}{}_{{#2}{#3}} }
\def\tilchris(#1-#2-#3){\tilde{{\mit \Gamma}}^{#1}{}_{{#2}{#3}}}
\def\hatchris(#1-#2-#3){\hat{{\mit \Gamma}}^{#1}{}_{{#2}{#3}}}

\DeclareFontFamily{OT1}{rsfs}{}
\DeclareFontShape{OT1}{rsfs}{m}{n}{ <-7> rsfs5 <7-10> rsfs7 <10-> rsfs10}{}

\newcommand{\Ric}{{\textrm{Ric}}}

\begin{document}

\begin{frontmatter}

\title{On the total mass of asymptotically hyperbolic manifolds\protect\thanksref{T1}}
\thankstext{T1}{Preprint UWThPh-2018-32}

\begin{aug}
    \author{\fnms{Hamed} \snm{Barzegar}\thanksref{t1}\ead[label=e1]{hamed.barzegar@univie.ac.at}}\thankstext{t1}{The author is supported by the Austrian Science Fund (FWF) project No. P29900-N27.},
    \address{Faculty of Physics and Erwin Schr\"odinger Institute\\
 University of Vienna \\Boltzmanngasse 5\\ A 1090 Wien, Austria\\
             \printead{e1}}
    \author{\fnms{Piotr T.} \snm{Chru\'{s}ciel}\thanksref{t2}
    \ead[label=e2]{piotr.chrusciel@univie.ac.at}%
    \ead[label=u2,url]{http://homepage.univie.ac.at/piotr.chrusciel}}
    \thankstext{t2}{The research of the author is supported by the Austrian Science Fund (FWF), Project  P 29517-N27, and
by the Polish National Center of Science (NCN) under grant 2016/21/B/ST1/00940.}
    \address{Faculty of Physics and Erwin Schr\"odinger Institute\\
 University of Vienna \\Boltzmanngasse 5\\ A 1090 Wien, Austria\\
             \printead{e2}\\
             \printead{u2}}
    \and
    \author{\fnms{Luc} \snm{Nguyen}
            \ead[label=e3]{luc.nguyen@maths.ox.ac.uk}}
    \address{Mathematical Institute and St Edmund Hall\\
    University of Oxford\\
    Andrew Wiles Building\\
    Radcliffe Observatory Quarter\\
    Woodstock Road, Oxford OX2 6GG, United Kingdom\\
             \printead{e3}}
\end{aug}


\begin{abstract}
Generalising a proof by Bartnik in the asymptotically Euclidean case, we give an elementary proof of positivity of the hyperbolic mass near the hyperbolic space.

It is a pleasure to dedicate this work to Robert Bartnik on the occasion of his 60th birthday.
\end{abstract}

\end{frontmatter}


\maketitle

\section{Introduction}

The question of positivity of total energy in general relativity has turned out to be a particularly challenging problem (cf. \cite{SchoenYau2017} and references therein), with several open questions remaining. It therefore appears of interest to provide simple proofs when available.

In his well-known paper on the mass of asymptotically Euclidean manifolds~\cite{Bartnik86}, Robert Bartnik gave an elementary proof of positivity of the ADM mass near the Euclidean metric. Inspired by his work, we establish a similar result for the hyperbolic mass near the hyperbolic metric. The argument turns out to be somewhat more involved and calculation-intensive.

Indeed, we provide an elementary proof of positivity of the hyperbolic mass, near the hyperbolic space, for metrics with scalar curvature bounded below by that of the hyperbolic space.
Namely,
ignoring an overall dimension-dependent constant, consider the usual definition (cf., e.g.~\cite{ChHerzlich}) of the   mass $m$   of a metric $g$ asymptotic to a metric $\backg$ with a static KID $V$ (see below for terminology):
\begin{multline}\label{2V18.11}
  m(V) = \lim_{R\to\infty} \int_{r=R}
   \big[
   V  g^{ m j}  g^{ i\ell }
    \left(
      \bcov{_ m} g_{ j\ell }
     -
     \bcov{_\ell } g_{ j m}
    \right)
    \\
    +
       (g^{ m j}  g^{ k i}
      - g^{ i  j}  g^{ km}
      )
     (g_{ j m} -\backg_{jm})\bcov{_ k}V
      \big]
      d\sigma_i
      \,.
\end{multline}
We prove the following:

\begin{theorem}
  \label{T16II18.1+}
  For  $n \ge 3$,
  let $(M,\backg)$ be  $\R^n$ equipped with the hyperbolic metric,
  \begin{equation}\label{26IV18.1}
    \backg = \frac{dr^2}{1+r^2} + r^2 d\Omega^2
    \,,
  \end{equation}
where $d\Omega^2$ is the canonical metric on the $(n-1)$-dimensional sphere $\mathbb{S}^{n-1}$.
Let
$(A^0 , \vec A)\in \R^{n+1}$ satisfy $|\vec A|:= \sqrt{(A^1)^2 + \ldots + (A^n)^2} \le A^0$ and set
\begin{equation}\label{17XII18.1}
  V= A^0 \sqrt{1+r^2} + \sum_i A^i x^i
   \,.
\end{equation}
Let $g$  be a  metric on $M$ asymptotic to $\backg$ with well-defined total mass $m$. There exists $\delta>0$ such that if
$$
  \|g-\backg\|_{L^\infty} + \|\bcov g\|_{L^\infty}<\delta
   \,,
$$
where $\bcov$ is the covariant derivative operator of $\backg$,
then $g$ can be put into the gauge
\begin{equation}\label{18XI18.1}
  \bcheckpsi^j:=\bcov{_ i}  g^{ij} - \frac{1}{2} g^{j k} \backg_{\ell m}\bcov_k g^{\ell m} =0
\end{equation}
in which we have
\begin{eqnarray}
 m(V)
 &\ge &
 \int_M \
  \Big[
  R - \overline R
   +
  \frac 1 {{8}n}
   |\bcov {g}|^2_{\backg}
    \Big]
     V \sqrtbg
    \label{23II18.1a}
\end{eqnarray}
where, in local coordinates, $\sqrtbg=\sqrt{\det \backg}\, d^{n}x$.
\end{theorem}

It follows clearly from \eqref{23II18.1a} that $m(V)\ge 0$ if
\begin{equation}
 R\ge \overline R
  \,.
   \label{17XII18.2}
\end{equation}
Equivalently, if we set
\begin{equation}\label{17XII18.4}
 m_0:= m(V=\sqrt{1+r^2})
 \,,
 \quad
  m_i := m (V=x^i)
 \,,
\end{equation}
then, under \eq{17XII18.2}, the vector $(m_\mu)$ is timelike future-pointing with respect to the Lorentzian quadratic form $m_0^2 -m_1^2\ldots - m_n^2$.
The inequality \eqref{17XII18.2}
holds of course for general relativistic initial data sets with vanishing trace of extrinsic curvature and with matter fields satisfying the dominant energy condition. Note that in vacuum, or in the presence of matter fields satisfying well behaved equations, under suitable further smallness assumptions on the extrinsic curvature of the initial data surface and on the matter fields, the condition of vanishing of the trace of the extrinsic curvature can be enforced by moving slightly the initial data hypersurface in space-time, after invoking the implicit-function theorem.

Theorem~\ref{T16II18.1+} is, essentially, a special case of Theorem~\ref{T16II18.1} below,
with the constants coming from \eqref{26IV18.8}.
At the heart of its proof lies the identity,   which we derive below
 and
which holds for any asymptotically hyperbolic background $(M,\backg)$ with a static KID $V$, under the usual conditions for existence of the  mass:%
\begin{eqnarray}
 m
 &
  \displaystyle
  = &
 \int_M \Big[  (R-\overline R) V
  + \Big(
 \quadratic{
   \frac{n+2}{8n}
   |\bcov \, \phi|^2_{\backg}
  +
   \frac{1}{4}
   |\bcov \hat{h}|^2_{\backg}}
   \nn
   \\
   &&
   \quadratic{-
     \frac{1}{2}
    \hat h^{i \ell} \hat h^{j m}  \adsR{_{\ell m i j}}
    -
    \frac{n+2}{2n} \phi \hat h^{i j} \adsR{_{ i j}}
     -
         \frac{n^2-4}{8n^2}  \lambda
      \phi^2
      }
\nn
  \\
    &&
    \gauge{
   -
    \frac{1}{2}
    \big(
       |\bcheckpsi|^2_{\backg}
       -
       \bcheckpsi^i\bcov_i \phi )
   }
 \Big)
 V
 +
  \Big(
    \tensor{h}{^k_i} \bcheckpsi^i
    +
    \frac 12
    \phi\bcheckpsi^k
   \Big)
   \bcov_k V
  \nn
\\
 & &
     +
    \Big(\herr
    +
    \hdhsqerr\Big)V
    \nn
 \\
 &&
 +
 \hdherr
  | \bcov  V|_{\backg}
 \Big]
  \sqrtbg
  \,;
   \label{12III18.1}
\end{eqnarray}
see \eq{28II18.1-}-\eq{20II18.1} for notation.
Throughout this work, the reader can assume that indices are raised and lowered using the background metric $\backg$.
We then use a weighted Poincar\'e inequality to control the non-obviously-positive terms in \eqref{12III18.1}.

The calculations leading to \eqref{12III18.1}, presented in Section~\ref{s14II18.1},
are vaguely reminiscent of those in~\cite{AbbottDeser}, but the relation of the formulae presented there to the hyperbolic mass is not clear.

We made an attempt to use similar ideas for perturbations of the Horowitz-Myers instantons~\cite{ConstableMyers,HorowitzMyers}, with only partial results so far~\cite{BCHMN}.

\begin{remark}
 \label{R22VIII19}
{\rm
The Birmingham-Kottler~\cite{Kottler,Birmingham} metrics with zero mass,
\begin{equation}\label{21III17.1}
 g= - \bigg( \frac{r^2}{\ell^2} +\kappa  \bigg) dt^2
   + \frac{dr^2} {\frac{r^2}{\ell^2} +\kappa  } + r^2 h_\kappa
 \,,
\ee
with constants  $\ell>0$ and   $\kappa\in\{0, \pm 1\}$,  where $({}^{n-1}\!N,h_\kappa)$ is an $(n-1)$-dimensional space form with Ricci scalar equal to $ (n-1)(n-2) \kappa  $, are space forms. Therefore all the \emph{calculations} here apply verbatim to the case of toroidal and hyperbolic conformal boundary at infinity for such metrics.
There are, however,  issues with the gauge, boundaries, and the weighted Poincar\'e inequality  which would need to be addressed to be able to obtain a positivity result:
\begin{enumerate}
  \item In the $\kappa=0$ case the associated manifold $(0,\infty)\times {} ^{n-1}\!N$ is complete with  one locally asymptotically hyperbolic end, where $r\to \infty$, and one cuspidal end, where $r\to 0$. Since the manifold is complete without boundary, the proof of existence of the gauge should go through for perturbations which vanish in the cuspidal end, but requires checking. We note that positivity of the mass in the spin case has been established in whole generality by Wang~\cite{Wang}, using a variation of Witten's proof, and in \cite{CGNP} in dimension $n\le 7$, but the non-spin higher dimensional case remains open.

  \item  In the   case  $\kappa =-1$ the manifold of interest is $[\ell,\infty)\times {}^{n-1}\!N$, where the boundary $\{\ell\}\times {}^{n-1}\!N$ satisfies a mean-curvature inequality. If the perturbations are \emph{not} supported away from the boundary there will be  terms arising from integration by parts  which are likely to destroy positivity, since in this case there exist well behaved solutions with negative mass.
\end{enumerate}
} %
\end{remark}

\section{Static KIDs}
 \label{s12VIII17.2}

Let $(M, \backg)$ be a smooth $n$-dimensional Riemannian manifold, $n \ge  2$ and let $V$ be a \emph{static KID} on $(M,\backg)$, i.e.\ a solution to
 \begin{equation}\label{12VIII1711-}
    \bcov_i \bcov_ j V = V \left(\bR_{ij} -   \frac {\bR}{n-1}  \backg_{ij} \right)
       \,.
 \end{equation}
When $\backg$ has constant scalar curvature, an equivalent form is
\begin{equation}\label{12VIII1711}
  \Delta_{\backg} V + \lambda V = 0
  \,,
   \quad
    \bcov_i \bcov_ j V = V (\bR_{ij} - \lambda \backg_{ij} )
\,,
\end{equation}
for some constant $\lambda  \in \R$.
 Here $\bR_{ij}$ denotes the Ricci tensor of the metric
$\backg$, $\bcov$ the Levi-Civita connection of $\backg$, and $\Delta_{\backg} =
 \bcov^k \bcov_k$
is the Laplacian of
$\backg$.

When $\lambda<0$,  rescaling $\backg$ by a constant factor if necessary, when the background metric has constant scalar curvature we can without loss of generality assume that
 $\lambda=-n$
  so that
$$
 \bR := \backg^{ij}\bR_{ij} = \lambda (n-1)=-n(n-1)
 \,.
$$
  \ptcheck{20XII17}

If $\backg$ is an Einstein  metric, namely  $\bR_{ij}$ proportional to $\backg_{ij}$, using this last scaling we obtain
\begin{equation}\label{20XI17.21}
  \bR_{ij} = - (n-1) \backg_{ij}
  \,,
   \quad
   \bcov_i \bcov_ j V = V  \backg_{ij}
    \,.
\end{equation}
This implies
\ptcheck{20XII17}
\begin{equation}\label{20XI17.22}
   \bcov_i( |\bcov  V |^2_{\backg} - V^2) = 0
    \,,
\end{equation}
where $|\cdot|_{\backg}$ denotes the norm of a tensor with respect to a  metric $\backg$.
In hyperbolic space, where the sectional curvatures are minus one, and when
$V$ takes the form \eqref{17XII18.1}  in the coordinate system of \eqref{26IV18.1},
we have
\begin{equation}\label{26IV18.7}
 |\bcov  V |^2_{\backg} - V^2 =|\vec A|^2 - (A^0)^2
 \,.
\end{equation}

\section{The theorem}
 \label{S26IV18.1}

It is convenient to introduce some notation:
\begin{eqnarray}
&
 h_{ij}:= g_{ij} - \backg_{ij}
 \,,
 &
  \label{28II18.1-}
\\
 &
 \psi^j:=\bcov{_ i}  g^{ij}
 \quad
 \Longleftrightarrow
 \quad
 g^{ij} \bcov_i h_{j\ell} =-g_{\ell j}  \psi^j
 \,,
  \label{28II18.1}
  &
\\
 &
 \phi := g^{ij} h_{ij}
 \quad
 \Longrightarrow
 \quad
  \overline \phi := \backg^{ij} h_{ij} = \phi +  O\left(|h |^2_{\backg} \right)
 \,.
  &
  \label{28II18.1+}
\end{eqnarray}
We will denote by $\check h$, respectively by $\hat h$, the $g$-trace-free, respectively the $\backg{}$-trace-free, part of $h$:
\begin{equation}\label{20II18.1}
 \check{h}_{ij}
 :=
 h_{ij} - \frac{1}{n}  {\phi} \, g_{ij}
 \,,
 \qquad
 \hat{h}_{ij}
 :=
 h_{ij} - \frac{1}{n} \overline{\phi} \, \backg_{ij}
  \,.
\end{equation}

In order to address the question of gauge-freedom, we will apply a diffeomorphism to $g$ so that
\begin{equation}\label{13II18.1a}
 \bcheckpsi^i:= \psi^i + \frac{1}{2} g^{i k} \bcov_k \phi
\end{equation}
vanishes. Note that the equation $\bcheckpsi^i = 0$ reduces to the harmonic-coordinates condition in the case of a flat background,
 where $\lambda=0$.

We claim the following:

\begin{theorem}
  \label{T16II18.1}
Let $(M,g)$ asymptote to an asymptotically hyperbolic space-form $(M,\backg)$ and let $V$ be a static KID of $(M,\backg)$.
Suppose that the usual decay conditions needed for a well-defined mass~\cite{ChHerzlich} are satisfied, namely, for large $r$, in the coordinate system of \eqref{26IV18.1},
\begin{multline}\label{16II18.12}
  h_{ij} = o(r^{-n/2})
  \,,
  \quad
  \bcov_k h_{ij} = O(r^{-n/2})
  \,,
   \\
  V = O(r)
  \,,
  \quad
  |\bcov_k h_{ij}|^2_{\backg}\, V \in L^1
  \,,
  \quad
  (R-\overline R) V \in L^1
  \,.
\end{multline}
There exists $\delta>0$ such that if
$$
  \|h\|_{L^\infty} + \|\bcov h\|_{L^\infty}<\delta
$$
and if $|dV|_{\backg}\le V$,
then
we have
\begin{eqnarray}
 m
 &\ge &
 \int_M \
  \Big[
  R - \overline R
   +
   \frac{n-2}{8n}
   |\bcov {h}|^2_{\backg}
    \Big]
     V \sqrtbg
     \nn
     \\
     &&
     -
    \frac{1}{2}
    \int_M
    \left((
       |\bcheckpsi|^2_{\backg}
       -
       \bcheckpsi^i\bcov_i \phi )
       V
       -
  \left(
    2 \tensor{h}{^k_i} \bcheckpsi^i
    +
    \phi\bcheckpsi^k
   \right)
   \bcov_k V
  \right)
    \sqrtbg
    \,.
    \label{23II18.1-}
\end{eqnarray}
\end{theorem}

A sharper bound can be found in \eqref{26IV18.8} below.

It follows from \eqref{26IV18.7} that  $|dV|_{\backg}\le V$ holds for static KIDs as in the statement of the theorem.  It is well known (cf.\ e.g., the proof of \cite[Theorem~4.5]{CDLS}; compare~\cite{GL,Lee:fredholm})
that the gauge $\bcheckpsi^k=0$ can always be realised  when $g$ is close enough to the hyperbolic metric $\backg{}$.
Hence Theorem~\ref{T16II18.1+} is indeed a corollary of Theorem~\ref{T16II18.1}.

\medskip

\proof\
 In Section~\ref{s14II18.1}  we prove the identity
\begin{equation}
 V \left( R - \bR \right) \rsqrtbg
 =
 \divergence{\overline{\mathcal{D}}}
  +
  \quadratic{\wmcQ }
 -
  \big(
    \tensor{h}{^k_i} \bcheckpsi^i
    +
    \frac 12
    \phi\bcheckpsi^k
   \big)
   \bcov_k V
     \,,
  \label{1II18.1a}
\end{equation}
where
\begin{eqnarray}
  \overline{\mathcal{D}}
  &
  :=
  &
  \bcov{_ i} \big[V  g^{ m j}  g^{ i\ell }
    \left(
      \bcov{_ m} h_{ j\ell }
     -
     \bcov{_\ell } h_{ j m}
    \right)
    \big]
    +
      \bcov{_ i}
      \left[
       (g^{ m j}  g^{ k i}
      - g^{ i  j}  g^{ km}
      )
     h_{ j m} \bcov{_ k}V
      \right]
\nn
 \\
  &&
  +
  \frac{1}{2}
  \bcov_i
\underbrace{   \left[
   V
    \backg_{k \ell}
      \left(
        g^{j k} \bcov_j g^{i \ell}
        -
        g^{i k} \bcov_j g^{j \ell}
      \right)
   \right]
   }_{(\diamond)}
   \nn
\\
 &&
   +
   \frac{1}{2}
     \bcov_i
  \left[
   \left(
  -3
   h^{i \ell} \tensor{h}{_\ell ^k}
  +
   g^{i k}
     |h|^2_{\backg}
   \right)
   \bcov_k V
  \right]
  \nn
\\
 &&
 { + \frac{1}{2}
      \bcov_i
       \left[
        \left(
         h^{k i} \phi
         +
         \frac{1}{4}
         \backg^{k i} \phi^2
        \right)
        \bcov_k V
       \right]
 \label{Sum16II18.11x}
   \rsqrtbg
   }
\end{eqnarray}
is the sum of all divergence terms
and $\wmcQ$ is the sum of all quadratic or higher order terms:
\begin{eqnarray}
\quadratic{
 \wmcQ
 }
 &
 \!\!\!=
 \Big(
 \!\!\!\!\!\!&%
 \quadratic{
  -
   \frac{n+2}{8n}
   |\bcov \, \phi|^2_{\backg}
   -
   \frac{1}{4}
   |\bcov \hat{h}|^2_{\backg}}
  \nn
   \\
   &&
   \quadratic{
   +
     \frac{1}{2}
    \hat h^{i \ell} \hat h^{j m}  \adsR{_{\ell m i j}}
    +
    \frac{n+2}{2n} \phi \hat h^{i j} \adsR{_{ i j}}
      +
         \frac{n^2-4}{8n^2} \lambda
      \phi^2
      }
\nn
  \\
    &&
    \gauge{
    +
    \frac{1}{2}
    \big(
       |\bcheckpsi|^2_{\backg}
       -
       \bcheckpsi^i\bcov_i \phi )
     }
     +
    \herr
    +
    \hdhsqerr
 \Big)
 V
 \nn
 \\
 &&
 +
 \hdherr
  | \bcov  V|_{\backg}
  \,.
   \label{Sum28II18.6}
\end{eqnarray}
Here the Riemann tensor can be replaced by the Weyl tensor, and the Ricci-tensor by its trace-free part.

We note that the term $(\diamond)$ in \eqref{Sum16II18.11x}
is quadratic in $(h,\bcov h)$:
\begin{eqnarray}
 \nonumber %
    \backg_{k \ell}
      \left(
        g^{j k} \bcov_j g^{i \ell}
        -
        g^{i k} \bcov_j g^{j \ell}
      \right)
 &
     =
   &
    (g_{k \ell} - h_{k\ell})
      \left(
        g^{j k} \bcov_j g^{i \ell}
        -
        g^{i k} \bcov_j g^{j \ell}
      \right)
\\
 &= &
      - h_{k\ell}
      \left(
        g^{j k} \bcov_j g^{i \ell}
        -
        g^{i k} \bcov_j g^{j \ell}
      \right)
      \,.
\end{eqnarray}
It is then easy to see that the integral of the divergence term $ \divergence{\overline{\mathcal{D}}}$  gives the total mass when integrated over the whole manifold, after taking into account the fact that the boundary conditions needed for a well-defined mass enforce a vanishing contribution of higher-than-linear terms in the boundary integral. This establishes \eqref{12III18.1}.

We specialise now to the space-form version \eqref{Sum28II18.6} of $\wmcQ $, which reads
\begin{eqnarray}
 \mathcal{Q}
 &=&
 \quadratic{
  \Big[-
   \frac{n+2}{8n}
   |\bcov \, \phi|^2_{\backg}
   -
   \frac{1}{4}
   |\bcov \hat{h}|^2_{\backg}
   +
     \frac{1}{2}
    |\hat{h}|^2_{\backg}
    -
    \frac{ n^2-4}{8n}
    \phi^2}
    \nn
    \\
    &&
    \quadratic{+
    \herr
    +
    O(|h|_{\backg} |\bcov h|^2_{\backg})
  \Big]
     V \rsqrtbg
     }
\nn
  \\
    &&
    \gauge{
    +
    \frac{1}{2}
    \big(
       |\bcheckpsi|^2_{\backg}
       -
       \bcheckpsi^i\bcov_i \phi )
       V
       \rsqrtbg
       }
          +
    \hdherr  |\bcov V|_{\backg} \rsqrtbg
    \,.
     \label{Sum26IV18.3}
\end{eqnarray}

In order to absorb the undifferentiated terms we use the weighted Poincar\'e inequality \eqref{14II18.3++} below, namely
\begin{eqnarray}
   \int  |  {\hat h}|_{\backg{}}^2 V
   \sqrtbg
   & \le  &
    \frac 1 {n}
    \int \Big[
     (|\bcov {\hat h}|_{\backg{}}^2-|\overline{{\fancyD}} {\hat h}|_{\backg{}}^2 - |\overline{\mathrm{div}}\, {\hat h} - {\hat h}_{dV}|_{\backg{}}^2
     )V
     \nn
     \\
     &&
     +
     \bcov_j({\hat h}_{ik}\bcov^iV {\hat h}^{jk})
     \Big]
     \sqrtbg
  \,.
  \label{Sum14II18.3++}
\end{eqnarray}
with $\fancyD$ defined in \eq{14II18.4}.
This leads to
\begin{eqnarray}
 \lefteqn{
\int \wmcQ  \sqrtbg
  \le
  \int
  \Big(
  \quadratic{
  -\Big[
   \frac{n+2}{8n}
   |\bcov \, \phi|^2_{\backg}
   +
   \frac{n-2}{4n}
   |\bcov \hat{h}|^2_{\backg}
    +
    \frac{ n^2-4}{8n}
    \phi^2
    }
    }
    &&
    \nn
    \\
    &&
    \quadratic{
    +  \frac 1 {2n}
    \big(
     |\overline{{\fancyD}} \hat h|_{\backg{}}^2
    +
      |\overline{\mathrm{div}}\, \hat h - \hat h_{dV}|_{\backg{}}^2
      \big)
    \Big]
     V \rsqrtbg
     }
\nn
  \\
    &&
     +
     \divergence{ \frac 1 {2n}
     \bcov_j(\hat h_{ik}\bcov^iV \hat h^{jk})
     }
    \gauge{
    +
    \frac{1}{2}
    \big(
       |\bcheckpsi|^2_{\backg}
       -
       \bcheckpsi^i\bcov_i \phi )
       V
       \rsqrtbg
       }
\nn
\\
 &&
     +
    \herr \absV
    +
    O(|h|_{\backg} |\bcov h|^2_{\backg}V)
          +
    \hdherr  |\bcov V|_{\backg}
     \Big) \sqrtbg
    \,.
\end{eqnarray}
Hence
\begin{eqnarray}
 m
 &\ge &
 \int_M \
  \Big[
  R - \overline R
  +
   \frac{n+2}{8n}
   |\bcov \, \phi|^2_{\backg}
   +
   \frac{n-2}{4n}
   |\bcov \hat{h}|^2_{\backg}
    +
    \frac{ n^2-4}{8n}
    \phi^2
    \nn
    \\
    &&
    +  \frac 1 {2n}
    \big(
     |\overline{{\fancyD}} \hat h|_{\backg{}}^2
    +
      |\overline{\mathrm{div}}\, \hat h - \hat h_{dV}|_{\backg{}}^2
      \big)
    \Big]
     V \sqrtbg
\nn
  \\
    &&
     -
    \int_M
    \Bigg(
    \gauge{
    \frac{1}{2}
    \big(
       |\bcheckpsi|^2_{\backg}
       -
       \bcheckpsi^i\bcov_i \phi )
       V
       \rsqrtbg
 -
  \left(
    \tensor{h}{^k_i} \bcheckpsi^i
    +
    \frac 12
    \phi\bcheckpsi^k
   \right)
   \bcov_k V
       }
\nn
\\
&&
       +
     \herr V
    +
    O(|h|_{\backg} |\bcov h|^2_{\backg}V)
          +
    \hdherr  |\bcov V|_{\backg}
     \Bigg) \sqrtbg
    \,.
     \phantom{xxxx}
     \label{26IV18.8}
\end{eqnarray}
It is now clear  that  we can choose   $|h|_{\backg} + |\bcov h|_{\backg}$ small enough so that
\eqref{23II18.1-} holds.
\qed

\section{The Ricci scalar of asymptotically anti de-Sitter spacetimes}
 \label{s14II18.1}

The aim of this section is to derive the the curvature identities \eqref{1II18.1a} and \eqref{Sum26IV18.3} needed in the proof of Theorem~\ref{T16II18.1}.

We consider the following metric
 \begin{equation} \label{12VIII17.1}
  g_{ij} = \backg{_{ij}} + h_{ij}\,,
 \end{equation}
where $\backg{_{ k i}}$ is an anti de-Sitter metric. If we denote the connection of the background metric by $\bcov{}$, we have the relation
 \begin{equation}
  D_ k \equiv \bcov{_ k} + \diffGamma{_ k}\,,
 \end{equation}
where $\diffGamma{_ k}$ is a $(1,2)$   tensor equal to $ \diffGamma{^{\cdot}_{\cdot k}} :=\tensor{\Gamma}{^{\cdot}_{\cdot k}}-\overline{\Gamma}\tensor{\vphantom{\Gamma}}{^{\cdot}_{\cdot k}}$.
For example, applying $D_ k$ on a vector component $v^ i$ we get
 \begin{equation}
  D_ k v^ i= \bcov{_ k} v^ i + \diffGamma{^ i_{ j k}} v^j\,,
 \end{equation}
where
 \ptcheck{14 VIII and rechecked 28 II}
\begin{eqnarray}\label{12VIII17.s}
 \diffGamma{^ i_{ j k}}
 &=&
  \frac{1}{2} g^{ i\ell }
  \left(
   \bcov{_j} g_{ k\ell } + \bcov{_ k} g_{\ell  j} - \bcov{_\ell } g_{ j k}
  \right)
  \nn
  \\
  &=&
  \frac{1}{2} g^{ i\ell }
  \left(
   \bcov{_ j} h_{ k\ell } + \bcov{_ k} h_{\ell  j} - \bcov{_\ell } h_{ j k}
  \right)\,.
 \end{eqnarray}
The Riemann tensors of the metrics $g_{ k i}$ and $\backg_{ k i}$ are related to each other via the following equation
 \ptcheck{28II18}
\begin{eqnarray}
\fullR{^ k_{ i m j}} = \adsR{^ k_{ i m j}} + \bcov{_m} \diffGamma{^ k_{ i j}} - \bcov{_ j} \diffGamma{^ k_{ i m}}
+ \diffGamma{^ k_{ m\ell }} \diffGamma{^\ell _{ i j}} -
\diffGamma{^ k_{ j\ell }} \diffGamma{^\ell _{ i m}}\,.
\end{eqnarray}
Contracting the first and third indices, one obtains
 \ptcheck{28II18}
 \begin{eqnarray}
  R_{ i j} = \adsR{_{ i j}} + \bcov{_ k} \diffGamma{^ k_{ i j}} - \bcov{_ j} \diffGamma{^ k_{ i k}}
+ \diffGamma{^ k_{ k\ell }} \diffGamma{^\ell _{ i j}} -
\diffGamma{^ k_{ j\ell }} \diffGamma{^\ell _{ i k}} \,.
\label{12VIII17.2}
 \end{eqnarray}
Inserting
 \ptcheck{up to here 15 VIII}
\begin{equation}
  \diffGamma{^ k_{ i k}} = \frac{1}{2} g^{ k\ell }
  \left(
   \bcov{_ i} h_{ k\ell }
   +
    \bcov{_ k} h_{ i\ell }
    -
    \bcov{_\ell } h_{ k i}
  \right)
  =
  \frac{1}{2} g^{ k\ell } \bcov{_ i} h_{ k\ell }
\end{equation}
into \eqref{12VIII17.2}, we obtain
 \ptcheck{28II18}
 \begin{eqnarray}
 \lefteqn{ R_{ i j}  = {\adsR{}}_{ i j}
 +
  \frac{1}{2} \Bigg[
  \bcov{_ k} g^{ k\ell }
   \left(
    \bcov{_ i} h_{ j\ell }
    +
    \bcov{_ j} h_{\ell i}
    -
    \bcov{_\ell } h_{ j i}
   \right)}
   \nn
 \\
 &&
 +
   g^{ k\ell }
   \left(
   \bcov{_ k} \bcov{_ i} h_{ j\ell }
    +
  \bcov{_ k}  \bcov{_ j} h_{\ell i}
    -
  \bcov{_ k}  \bcov{_\ell } h_{ j i}
   \right)
   -
   \bcov{_ j} g^{ k\ell } \bcov{_ i} h_{ k\ell }
   \nn
   \\
   &&
   -
    g^{ k\ell } \bcov{_ j} \bcov{_ i} h_{ k\ell }
    +
    \frac{1}{2} g^{ k p} g^{\ell q} \bcov{_\ell } h_{ p k}
    \left(
     \bcov{_ i} h_{ j q}
     +
      \bcov{_ j} h_{ i q}
     -
     \bcov{_ q} h_{ i j}
    \right)
    \nn
    \\
    &&
    -
    \frac{1}{2} g^{ k p} g^{\ell q}
   \left(
     \bcov{_ j} h_{\ell p}
     +
      \bcov{_\ell } h_{ p j}
     -
     \bcov{_ p} h_{ j\ell }
    \right)
     \left(
     \bcov{_ i} h_{ k q}
     +
      \bcov{_ k} h_{ i q}
     -
     \bcov{_ q} h_{ k i}
    \right)
    \Bigg]
   .
   \phantom{xx}
    \label{20VIII17.1}
 \end{eqnarray}
And the Ricci scalar reads
 \ptcheck{15VIII and rechecked 28 II}
 \begin{eqnarray}
  R &=& g^{ i j} R_{ i j}
  \nn
  \\
  &=&
   \adsR{_{ij}}  g^{ i j}
  +
   \frac{1}{2} g^{ i j} \Bigg[
    \bcov{_ k} g^{ k\ell }
    \left(
     2  \bcov{_ i} h_{ j\ell }
     -
      \bcov{_\ell } h_{ j i}
    \right)
   \nn
    \\
    &&
    +
    2 g^{ k\ell }
    \left(
     \bcov{_ k}  \bcov{_ i} h_{ j\ell }
     -
       \bcov{_ k} \bcov{_\ell } h_{ j i}
    \right)
    \nn
    \\
    &&
    -
    \bcov{_ j} g^{ k\ell } \bcov{_ i} h_{ k\ell }
    +
     \frac{1}{2}
     g^{ k p} g^{\ell q} \bcov{_\ell } h_{ p k}
    \left(
     2 \bcov{_ i} h_{ j q}
     -
     \bcov{_ q} h_{ i j}
    \right)
    \nn
    \\
    &&
    -
     \frac{1}{2} g^{ k p} g^{\ell q}
   \left(
     \bcov{_ j} h_{\ell p}
     +
      \bcov{_\ell } h_{ p j}
     -
     \bcov{_ p} h_{ j\ell }
    \right)
     \left(
     \bcov{_ i} h_{ k q}
     +
      \bcov{_ k} h_{ i q}
     -
     \bcov{_ q} h_{ k i}
    \right) \Bigg]\,.
    \label{12VIII17.3}
     \nn
    \\
 \end{eqnarray}
Using
\ptcheck{28II18}
\begin{eqnarray}
\lefteqn{
  \frac{1}{2} g^{ i j} g^{ k p} g^{\ell q}
   \left(
     \bcov{_ j} h_{\ell p}
     +
      \bcov{_\ell } h_{ p j}
     -
     \bcov{_ p} h_{ j\ell }
    \right)
     \left(
     \bcov{_ i} h_{ k q}
     +
      \bcov{_ k} h_{ i q}
     -
     \bcov{_ q} h_{ k i}
    \right)
    }
    &&
    \nn
    \\
    &&
     \phantom{xxxxx}
     =
    \frac{1}{2} g^{ i j} g^{ k p} g^{\ell q}
    \bcov{_ p} h_{ j\ell }
    \left(
     2 \bcov{_ q} h_{ k i}
     -
     \bcov{_ k} h_{ i q}
    \right)\,,
     \label{19XII17.21}
\end{eqnarray}
this can be rewritten as
 \ptcheck{28II18}
 \begin{eqnarray}
  R &=&
   \adsR{_{ij}}  g^{ i j}
  +
   g^{ i j}  g^{ k\ell }
    \left(
     \bcov{_ k}  \bcov{_ i} h_{ j\ell }
     -
       \bcov{_ k} \bcov{_\ell } h_{ j i}
    \right)
   \nn
    \\
    &&
  +
   \frac{1}{2} g^{ i j}
   \Big[
    \bcov{_ k} g^{ k\ell }
    \left(
     2  \bcov{_ i} h_{ j\ell }
     -
      \bcov{_\ell } h_{ j i}
    \right)
    \nn
    \\
    &&
    -
    \bcov{_ j} g^{ k\ell } \bcov{_ i} h_{ k\ell }
    +
     \frac{1}{2}
     g^{ k p} g^{\ell q}
      \Big[
     \bcov{_\ell } h_{ p k}
    \left(
     2 \bcov{_ i} h_{ j q}
     -
     \bcov{_ q} h_{ i j}
    \right)
    \nn
 \\
  &&
    -
    \bcov{_ p} h_{ j\ell }
    \left(
     2 \bcov{_ q} h_{ k i}
     -
     \bcov{_ k} h_{ i q}
    \right)
    \Big]
    \Big]\,.
    \label{17XII17.3+}
 \end{eqnarray}

In order to isolate the contribution of the mass we group all second-derivative terms in \eqref{12VIII17.3} in a divergence  \emph{with respect to the background metric} (similar to~\cite{ChHerzlich}, except that there the divergence was taken with respect to the physical metric):
 \ptcheck{28II18}
 \begin{eqnarray}
  R &=&
  \adsR{_{ij}} g^{ i j}
  +
   \bcov{_ k} \big[ g^{ i j}  g^{ k\ell }
    \left(
      \bcov{_ i} h_{ j\ell }
     -
     \bcov{_\ell } h_{ j i}
    \right)
    \big]
   \nn
    \\
    &&
     -
   \bcov{_ k} \left( g^{ i j}  g^{ k\ell }\right)
    \left(
      \bcov{_ i} h_{ j\ell }
     -
     \bcov{_\ell } h_{ j i}
    \right)
    \nn
    \\
    &&
     +
   \frac{1}{2} g^{ i j}
   \Big[
    \bcov{_ k} g^{ k\ell }
    \left(
     2  \bcov{_ i} h_{ j\ell }
     -
      \bcov{_\ell } h_{ j i}
    \right)
    -
    \bcov{_ j} g^{ k\ell } \bcov{_ i} h_{ k\ell }
    \nn
    \\
    &&
    +
     \frac{1}{2}
     g^{ k p} g^{\ell q}
      \left[
     \bcov{_\ell } h_{ p k}
    \left(
     2 \bcov{_ i} h_{ j q}
     -
     \bcov{_ q} h_{ i j}
    \right)
    -
    \bcov{_ p} h_{ j\ell }
    \left(
     2 \bcov{_ q} h_{ k i}
     -
     \bcov{_ k} h_{ i q}
    \right)
    \right]
    \Big]\,.
    \label{12VIII17.3+}
     \nn
    \\
 \end{eqnarray}
Note that
$$
 0 = \bcov{} _ j \delta^ k_ i = \bcov{}_ j (g^{ k p} g_{ p  i})
  = g_{ p  i} \bcov{}_ j  g^{ k p}+ g^{ k p} \bcov{}_ j  g_{ p  i}
  \,,
$$
equivalently
\begin{equation}\label{22XII17.1}
  \bcov{}_ j  g^{ k p}= -
    g^{\ell  k} g^{ i p} \bcov{}_ j  g_{ i\ell }
    =-
    g^{\ell  k} g^{ i p} \bcov{}_ j  h_{ i\ell }
  \,.
\end{equation}
This allows us to rewrite \eq{12VIII17.3+} as
 \ptcheck{28II18}
 \begin{eqnarray}
  R &=&
  \adsR{_{ij}} g^{ i j}
  +
   \bcov{_ k} \big[ g^{ i j}  g^{ k\ell }
    \left(
      \bcov{_ i} h_{ j\ell }
     -
     \bcov{_\ell } h_{ j i}
    \right)
    \big]
     -
   \bcov{_ k} \left( g^{ i j}  g^{ k\ell }\right)
    \left(
      \bcov{_ i} h_{ j\ell }
     -
     \bcov{_\ell } h_{ j i}
    \right)
     \nonumber
   \\
  &&
     +
     \frac{1}{2} g^{ i j}
   \Big[
    -g^{ k p} g^{\ell q} \bcov{_ k} h_{ p q}
    \left(
     2  \bcov{_ i} h_{ j\ell }
     -
      \bcov{_\ell } h_{ j i}
    \right)
    +
    g^{ k p} g^{\ell q}\bcov{_ j} h_{ p q} \bcov{_ i} h_{ k\ell }
    \nn
    \\
    &&
    +
     \frac{1}{2}
     g^{ k p} g^{\ell q}
      \left[
     \bcov{_\ell } h_{ p k}
    \left(
     2 \bcov{_ i} h_{ j q}
     -
     \bcov{_ q} h_{ i j}
    \right)
    -
    \bcov{_ p} h_{ j\ell }
    \left(
     2 \bcov{_ q} h_{ k i}
     -
     \bcov{_ k} h_{ i q}
    \right)
    \right]
    \Big]
  \nn
  \\
  &=&
   \adsR{_{ij}} g^{ i j}
  +
   \bcov{_ k} \big[ g^{ i j}  g^{ k\ell }
    \left(
      \bcov{_ i} h_{ j\ell }
     -
     \bcov{_\ell } h_{ j i}
    \right)
    \big]
  \nn
    \\
    &&
    +
     \left(
  g^{ k\ell } g^{ i p} g^{ j q} \bcov{_ k} h_{ p q}
  +
  g^{ i j} g^{ k p} g^{\ell q} \bcov{_ k} h_{ p q}
  \right)
  \left(
      \bcov{_ i} h_{ j\ell }
     -
     \bcov{_\ell } h_{ j i}
    \right)
     \nonumber
   \\
  &&
     +
     \frac{1}{2} g^{ i j} g^{ k p} g^{\ell q}
   \Big[
    - \bcov{_ k} h_{ p q}
    \left(
     2  \bcov{_ i} h_{ j\ell }
     -
      \bcov{_\ell } h_{ j i}
    \right)
    +
    \bcov{_ j} h_{ p q} \bcov{_ i} h_{ k\ell }
    \nn
    \\
    &&
    +
     \frac{1}{2}
      \left[
     \bcov{_\ell } h_{ p k}
    \left(
     2 \bcov{_ i} h_{ j q}
     -
     \bcov{_ q} h_{ i j}
    \right)
    -
    \bcov{_ p} h_{ j\ell }
    \left(
     2 \bcov{_ q} h_{ k i}
     -
     \bcov{_ k} h_{ i q}
    \right)
    \right]
    \Big]
     \,.
\end{eqnarray}

After some simplifications one gets
\ptcheck{28II18}
\begin{equation}\label{12VIII17.3+b}
 R
 =
 \adsR{_{ij}} g^{ i j}
  +
   \bcov{_ k} \big[ g^{ i j}  g^{ k\ell }
    \left(
      \bcov{_ i} h_{ j\ell }
    -
    \bcov{_\ell } h_{ j i}
    \right)
    \big]
    +
    Q
    \,,
\end{equation}
where
 \ptcheck{15 VIII  the result with luc up to here (but not all intermediate formulae) and rechecked by Hamed on 1 II 18 and rechecked including all intermediate formulae now}
 \begin{eqnarray}
  \label{17XII17.1}
 Q &:= &
 \underbrace{\frac 14   g^{ i j} g^{ k p} g^{\ell q}
  \Big(
  2
    \bcov{_ p} h_{ j\ell }  \bcov{_ q} h_{ k i}}_{=:Q_1}
   \quadratic{
     -
     \bcov{_\ell } h_{ k p}   \bcov{_ q} h_{ i j}
     -
     \bcov{_ i} h_{ p q}   \bcov{_ j} h_{ k\ell }
     }
  \Big)
  \,.
  \phantom{xxx}
 \end{eqnarray}
%

We note that
%
\begin{equation}
 g^{ij}
 =
 \backg^{ij} - h^{ij} + \chi^{ij}
 \,,
 \label{2V18.1}
\end{equation}
where
\begin{equation}
  h^{i}{}_{\ell} = \backg^{i k}  h_{k \ell}
  \,,
  \quad
  h^{i j} = \backg^{i k} \backg^{j \ell } h_{k \ell}
  \,,
\end{equation}
and
\begin{equation}
 \chi^{ij} :=
    \backg^{ik}\backg^{j\ell}\backg^{mn}h_{km}h_{n\ell}
       +O(|h |^3 _{\backg}) =
       O(|h|_{\backg}^2)
 \,.
  \label{2V18.1+}
\end{equation}

In the notation of \eqref{28II18.1}-\eqref{13II18.1a}, the identity~\eqref{12VIII17.3+b} becomes
\begin{equation}\label{28II18.5}
- \frac 12 \bcov_k \left( g^{kl}D_l \phi
 \right)  =  R -
 \adsR{ }
  +
 \adsR{_{ij} } h^{ij}
 -
  \underbrace{
   \bcov{_ k} \left(   g^{ k\ell } h_{ j i}
    \bcov{_\ell } g^{ij} - \bcheckpsi ^k
    \right)
     \underbrace{-Q}_{O(|h|^2 + |\bcov h|^2)}
    }_{\textrm{``higher order terms''}}
  \! \! \! \!\!\!\!\!\!\! .
\end{equation}

 If both $g$ and $\backg$ satisfy the vacuum scalar constraint equation, so that $R=\adsR{}$, and in the gauge $\bcheckpsi ^i =0$,
\eqref{28II18.5} takes the form
\begin{eqnarray}
\lefteqn{
- \frac 12 \bcov_k \left( g^{kl}D_l \phi
 \right) -
 \frac {\adsR{  } } n   \phi
 }
   &&
   \nn
   \\
   &=&
 \adsR{_{ij} } \hat h^{ij}
  \underbrace{
 -
 \frac {\adsR{  } } n (  \phi -
\overline  \phi)
 - \bcov{_ k} \left(   g^{ k\ell } h_{ j i}
    \bcov{_\ell } g^{ij}
    \right)
    + O(|h|^2 + |\bcov h|^2)
    }_{\textrm{``higher order terms''}}
    \,,
    \phantom{xxx}
    \label{28II18.4}
\end{eqnarray}
which becomes an elliptic equation for $\phi$ when all ``higher order terms'' are thought to be negligible. Note that when $\backg$ is a space-form the linear term at the right-hand side vanishes, which implies that $\phi$ itself is higher order.
However, this is not true in general, in particular one cannot assume that $\phi=0$ for general perturbations of e.g.\ the Horowitz-Myers metrics.

We return to the calculation of the mass. Let $V$ be a static KID as in Section~\ref{s12VIII17.2}.
Multiplying \eqref{12VIII17.3+b} by $V\rsqrtbg $ we obtain
\begin{eqnarray}
 \nonumber
 V R \rsqrtbg  &=&
 V \Big(
  \adsR{_{ij}} g^{ i j}
  +
   \bcov{_ k} \big[ g^{ i j}  g^{ k\ell }
    \left(
      \bcov{_ i} h_{ j\ell }
     -
     \bcov{_\ell } h_{ j i}
    \right)
    \big]
     + Q
     \Big) \rsqrtbg
     \nonumber
      \\
   &=&
 V \Big(
  \adsR{_{ij}} g^{ i j}
     + Q
     \Big) \rsqrtbg
      + {{\rrho}}
      \,,
   \label{14VII17.31}
\end{eqnarray}
where
\begin{eqnarray}
 \nonumber
  {{\rrho}} &:=&
 V
   \bcov{_ k} \big[ g^{ i j}  g^{ k\ell }
    \left(
      \bcov{_ i} h_{ j\ell }
     -
     \bcov{_\ell } h_{ j i}
    \right)
    \big]
    \rsqrtbg
     \nonumber
\\
  &=&
    \divergence{
    \bcov{_ k} \big[V    g^{ i j}  g^{ k\ell }
    \left(
      \bcov{_ i} h_{ j\ell }
     -
     \bcov{_\ell } h_{ j i}
    \right)
    \big] \rsqrtbg
      }
   \underbrace{ -
    g^{ i j}  g^{ k\ell }
    \left(
      \bcov{_ i} h_{ j\ell }
     -
     \bcov{_\ell } h_{ j i}
    \right)
      \bcov{_ k}V
     \rsqrtbg }_{=:*}
      \,.
       \phantom{xxxx}
      \label{14VII17.33}
\end{eqnarray}
%
Then
\begin{eqnarray}
 \nonumber
  *
 & = &
    \Big( -
    \bcov{_ i} \big( g^{ i j}  g^{ k\ell }
      h_{ j\ell } \bcov{_ k}V
      \big)
      +
     h_{ j\ell } \bcov{_ i} (
 g^{ i j}  g^{ k\ell }\bcov{_ k}V )
 \nonumber
\\
 &&
  +
     \bcov{_\ell }\big(
      g^{ i j}  g^{ k\ell }
     h_{ j i} \bcov{_ k}V \big)
     -
    h_{ j i}\bcov{_\ell }\big(  g^{ i j}  g^{ k\ell }
     \bcov{_ k}V\big)
   \Big)
     \rsqrtbg
  \nonumber
\\
 & = &
    \Big(
    \divergence{
      \bcov{_ \ell}\big(
       (g^{ i j}  g^{ k \ell}
      - g^{ \ell  j}  g^{ k i}
      )
     h_{ j i} \bcov{_ k}V
      \big)
      }
      +
      \care{used in \eqref{19XII17.3}}{
      h_{ j \ell}
     (
 g^{ i j}  g^{ k \ell}  -
       g^{\ell j}  g^{ k i}
       )
     \bcov{_ i}\bcov{_ k}V
     }
 \nonumber
\\
 &&  +
    \care{first term in \eqref{9II18.2}}{
     h_{ j \ell} \bcov{_ i} \big(
 g^{ i j}  g^{ k \ell}
 - g^{ \ell j}  g^{ k i}\big)
     \bcov{_ k}V
     }
   \Big)
     \rsqrtbg
      \,.
         \label{15VII17.11}
\end{eqnarray}

The last two terms in \eqref{17XII17.1} are manifestly negative, which is the desired sign for our purposes. The part $Q_1$ of $Q$ requires further manipulations, as follows:
%
\begin{eqnarray}
  \nn
 V Q_1\rsqrtbg
 &= &
  \frac 12   Vg^{ i j} g^{ k p} g^{\ell q}
    \bcov{_ p} h_{ j\ell }  \bcov{_ q} h_{ k i}
 \rsqrtbg
\\
 &=&
  \frac 12   V \backg^{i j}
  \bcov{_ k} \tensor{h}{ _j^\ell }  \bcov{_ \ell} \tensor{h}{ ^k _i}
 \rsqrtbg
  +
 \hdhsqerr   \rsqrtbg V
\nn
 \\
 &=&
  \frac 12   V
  \backg_{k \ell}
  \bcov_i g^{j k} \bcov_j g^{i \ell}
   \rsqrtbg
  +
  \hdhsqerr  \rsqrtbg V
\nn
 \\
  &=&
  \frac 12
    V
   \Big\{
  \bcov_i
   \left[
    \backg_{k \ell}
      \left(
        g^{j k} \bcov_j g^{i \ell}
        -
        g^{i k} \bcov_j g^{j \ell}
      \right)
   \right]
   +
   \backg_{k \ell}
   \bcov_i g^{i k} \bcov_j g^{j \ell}
   \nn
   \\
   &&
   -
   \backg_{k \ell} g^{i k}
    \left(
     \bR_{m i} g^{m \ell}
     -
     \adsR{^\ell _{m i j}} g^{j m}
    \right)
    \rsqrtbg
   +
   \hdhsqerr \rsqrtbg
    \Big\}
   \,.
   \label{19II18.1}
\end{eqnarray}
%
In the notation of \eqref{2V18.1},
Equation~\eqref{19II18.1} becomes
\begin{eqnarray}
 V Q_1
 &=&
 \frac 12
    V
   \Big\{
  \bcov_i
   \left[
    \backg_{k \ell}
      \left(
        g^{j k} \bcov_j g^{i \ell}
        -
        g^{i k} \bcov_j g^{j \ell}
      \right)
   \right]
   +
   \backg_{k \ell}
   \bcov_i g^{i k} \bcov_j g^{j \ell}
  \nn
  \\
   &&
   -
   \chi^{i j} \adsR{_{i j}}
   +
   \tensor{h}{^i _\ell} h^{j m} \adsR{^\ell_{m i j}}
    \rsqrtbg
   +
   \herr
   +
   \hdhsqerr \rsqrtbg
    \Big\}
\nn
  \\
  &=&
   \frac 12
     \Big\{
      \divergence{
  \bcov_i
   \left[
   V
    \backg_{k \ell}
      \left(
        g^{j k} \bcov_j g^{i \ell}
        -
        g^{i k} \bcov_j g^{j \ell}
      \right)
   \right]
   }
   \nn
  \\
   &&
   \care{second term in \eqref{9II18.2}}{
   -
   \backg_{k \ell}
      \left(
        g^{j k} \bcov_j g^{i \ell}
        -
        g^{i k} \bcov_j g^{j \ell}
      \right)
      \bcov_i V
        }
\nn
  \\
  &&
      +
      \left(
      |\psi|^2_{\backg}
   -
   \chi^{i j} \adsR{_{i j}}
   +
   \tensor{h}{^i _\ell} h^{j m} \adsR{^\ell_{m i j}}
   \right)
   V
   \Big\}
   \rsqrtbg
  \nn
  \\
   &&
   +
   \herr V
   +
   \hdhsqerr V \rsqrtbg
   \,.
   \label{19II18.2}
\end{eqnarray}
In the special case where $\backg$ is a (suitably normalised) hyperbolic space-form
we have
\begin{equation}
 \adsR{^i_{j k \ell}}
 =
  \frac{\adsR{}}{n(n-1)}
 \left(
  \delta^i_k \backg_{j \ell}
  -
  \delta^i_\ell \backg_{j k}
 \right)
 =
 -
  \left(
  \delta^i_k \backg_{j \ell}
  -
  \delta^i_\ell \backg_{j k}
 \right)
 \,,
\end{equation}
and the relations in \eqref{20XI17.21} are satisfied.
In this case \eqref{19II18.2} becomes
 \ptcheck{28 II 18 the orange terms again}
\begin{eqnarray}
V Q_1
  &=&
     \frac 12
     \Big\{
      \divergence{
  \bcov_i
   \left[
   V
    \backg_{k \ell}
      \left(
        g^{j k} \bcov_j g^{i \ell}
        -
        g^{i k} \bcov_j g^{j \ell}
      \right)
   \right]
   }
 \nn
  \\
   &&
   -
   \backg_{k \ell}
      \left(
        g^{j k} \bcov_j g^{i \ell}
        -
        g^{i k} \bcov_j g^{j \ell}
      \right)
      \bcov_i V
\nn
  \\
  &&
   \quadratic{
   +
   \left(
   |\psi|^2_{\backg}
   -
   \phi^2
   +
   n
   |h|^2_g
   \right)
   V
   }
   \Big\}
   \rsqrtbg
   +
   \herr V
   +
   \hdhsqerr V  \rsqrtbg
   .
   \phantom{xxxx}
   \label{8II18.1}
\end{eqnarray}

In order to simplify the expressions derived so far we consider similar terms separately:

\begin{enumerate}

\item
We wish to add the second term of \eqref{19II18.2}  and the third term of \eqref{15VII17.11}:
%

\ptcheck{14II18}
%
\begin{eqnarray}
\lefteqn{
h_{ j \ell} \bcov{_ i}
      (
 g^{ i j}  g^{ k \ell}
 - g^{ \ell j}  g^{ k i})
     \bcov{_ k}V
     \rsqrtbg
  -
  \frac 12
   \backg_{k \ell}
      \left(
        g^{j k} \bcov_j g^{i \ell}
        -
        g^{i k} \bcov_j g^{j \ell}
      \right)
      \bcov_i V
        \rsqrtbg
        }
      &
\nn
 \\
    &=&
     h_{j \ell}
      \left(
       \psi^j g^{k \ell}
       +
       g^{i j} \bcov_i g^{k \ell}
       -
       \bcov_i g^{\ell j} g^{k i}
       -
       g^{\ell j} \psi^k
      \right)
       \bcov_k V \rsqrtbg
\nn
 \\
   &&
   +
   \frac{1}{2}
    \left(
     h_{k \ell} g^{j k} \bcov_j g^{i \ell}
     -
     h_{k \ell} g^{i k} \psi^\ell
    \right)
    \bcov_i V \rsqrtbg
\nn
  \\
    &=&
    \frac{1}{2}
    \Big[
     \tensor{h}{^k_j } \psi^j
     +
     3
     \tensor{h}{^i_ \ell} \bcov_i g^{k \ell}
     +
     g^{k i} \bcov_i |h|^2_{\backg}
     -
     2\phi
      \psi^k
      +
      \hdherr
   \Big]
   \bcov_k V \rsqrtbg
\nn
  \\
   &=&
  \underbrace{
   \frac{1}{2}
     \Big[
     \gauge{
      -
      2 \tensor{h}{^k_i} \psi^i
        -
      2 \phi
         \psi^k
         }
        }_{=:\mathcal{ A}^k  \,,\ \mbox{\rm \scriptsize taken care of in \eqref{26IV18.9}}}
       \underbrace{
       -
      3 \bcov_i \left( h^{i \ell} \tensor{h}{_\ell ^k} \right)
      +
      g^{ki} \bcov_i |h|^2_{\backg}
      }_{=:\mathcal{P}^k\,,\ \mbox{\rm \scriptsize taken care of in \eqref{25I18.1}}}
  \nn
  \\
   &&
      +
      \hdherr
     \Big]
      \bcov{_ k}V  \rsqrtbg
      \,,
   \phantom{xxxxx}
   \label{9II18.2}
\end{eqnarray}
where we used
\ptcheck{9I18}
\begin{eqnarray}
 3 \tensor{h}{^i _\ell} \bcov_i g^{k \ell}
 &=&
 -3 \tensor{h}{^i _\ell} \bcov_i h^{k \ell}
  +
 \hdherr
 \nn
 \\
 &=&
 -
 3
 \bcov_i \left( h^{i \ell} \tensor{h}{_\ell ^k}  \right)
 +
 3
 \bcov_i h^{i \ell}  \tensor{h}{_\ell ^k}
 +
 \hdherr
 \nn
 \\
 &=&
  -
 3
 \bcov_i \left( h^{i \ell} \tensor{h}{_\ell ^k}  \right)
 -
 \gauge{
 3
 \tensor{h}{ ^k _i} \psi^i }
 +
 \hdherr
 \,.
\end{eqnarray}

We may rewrite the terms indicated by $\mathcal{P}$ in terms of total divergences as follows
\ptcheck{26II18, and Ricci term rechecked 28 II 18}
\begin{eqnarray}
 \lefteqn{
   \frac{1}{2}
 \mathcal{P}^k \bcov_k V \rsqrtbg
 =
  \frac{1}{2}
  \bcov_i
  \left[
   \left(
  -3
   h^{i \ell} \tensor{h}{_\ell ^k}
  +
   g^{i k}
     |h|^2_{\backg}
   \right)
   \bcov_k V
  \right]
 \rsqrtbg
 }
 &&
  \nn
\\
  &&
  +
   \frac{1}{2}
  \big(
   3
  h^{i \ell} \tensor{h}{_\ell ^k}
  -
  g^{ik}
   \sqrdof{h}
  \big)
  V
   \left(
    \adsR{_{i k}}
    -
    \lambda  \backg_{i k}
   \right)
  \rsqrtbg
  +
  \hdherr
  | \bcov  V| \rsqrtbg
\nn
  \\
  &=&
  \divergence{
  \frac{1}{2}
  \bcov_i
  \left[
   \left(
  -3
   h^{i \ell} \tensor{h}{_\ell ^k}
  +
   g^{i k}
     \sqrdof{h}
   \right)
   \bcov_k V
  \right]
 \rsqrtbg
 }
  \quadratic{
  +
   \frac{1}{2}
   \Big\{
     3
     h^{i \ell} \tensor{h}{_\ell ^k} \adsR{_{i k}}
     }
  \nn
\\
  &&
  \quadratic{
     -
     \left[
       \adsR{}
       +
       \lambda \left( 3-n \right)
     \right]
     \sqrdof{h}
     +
     \herr
   \Big\}  V
   +
     \hdherr
  | \bcov  V| \rsqrtbg
  }
  \,.
  \nn
  \\
  \label{25I18.1}
\end{eqnarray}
When $\backg$ is Einstein, the result simplifies to:
\ptcheck{31I18 and h square terms rechecked 28 II}
\begin{eqnarray}
 \lefteqn{
 \frac{1}{2}
 \mathcal{P}^k \bcov_k V \rsqrtbg
   =
  \divergence{
    \frac{1}{2}
     \bcov_i
  \left[
   \left(
  -3
   h^{i \ell} \tensor{h}{_\ell ^k}
  +
   g^{i k}
     |h|^2_{\backg}
   \right)
   \bcov_k V
  \right]
  \rsqrtbg
  }
  }
  &&
   \nn
\\
  &&
  +
  \quadratic{
   \frac{1}{2}
  \big[
   (3-n)
  |h|^2_{\backg}
  +
  \herr
  \big]
  V
  \rsqrtbg
  +
  \hdherr
  | \bcov  V| \rsqrtbg
  }
  \,.
  \label{25I18.1-}
\end{eqnarray}

Returning to the general case, in the notation of \eqref{9II18.2}  we find
\ptcheck{26II18}
\begin{eqnarray}
\nn
\lefteqn{
 \mathcal{ A}^k \bcov_k V
}
 &&
 \\
 &
  \equiv
  &
  -
  \left(
    \tensor{h}{^k_i} \psi^i
    +
    \phi\psi^k
   \right)
   \bcov_k V
   \nn
\\
  &= &
  -
  \left(
    \tensor{h}{^k_i} \bcheckpsi^i
    +
    \phi\bcheckpsi^k
   \right)
   \bcov_k V
   +
   \frac{1}{2}
    \left(
       \tensor{h}{^k_i} g^{i \ell} \bcov_\ell \phi
       +
       \phi g^{k \ell} \bcov_\ell \phi
    \right)
    \bcov_k V
\nn
  \\
    &=&
  -
  \left(
    \tensor{h}{^k_i} \bcheckpsi^i
    +
    \phi\bcheckpsi^k
   \right)
   \bcov_k V
 \nn
 \\
 &&
   +
   \frac{1}{2}
     \left[
      h^{k \ell} \bcov_\ell \phi
      +
      \frac{1}{2}
      \backg^{k \ell} \bcov_\ell \phi^2
      +
      \hdherr
     \right]
     \bcov_k V
\nn
  \\
    &=&
  -
  \left(
    \tensor{h}{^k_i} \bcheckpsi^i
    +
    \phi\bcheckpsi^k
   \right)
   \bcov_k V
   +
    \frac{1}{2}
    \Bigg\{
     \bcov_\ell
       \left[
        \left(
         h^{k \ell} \phi
         +
         \frac{1}{2}
         \backg^{k \ell} \phi^2
        \right)
        \bcov_k V
       \right]
      \nn
\\
 &&
       +
      \underbrace{
       (\psi^k -\bcheckpsi^k) \phi \bcov_k V
      }_{= -\frac{1}{4}
                 \left[
                   \bcov_\ell
                      \left(
                        \backg^{k \ell} \phi^2 \bcov_k V
                      \right)
                      -
                        \backg^{k \ell} \phi^2 \bcov_\ell \bcov_k V
                        +
                        \hdherr |\bcov V|_{\backg}
                 \right]
          }
          + \bcheckpsi^k \phi \bcov_k V
\nn
 \\
    &&
       -
       \big(h^{k \ell } \phi
       +
       \frac{1}{2}
       \backg^{k \ell} \phi^2
       \big)
       \bcov_k \bcov_\ell V
       +
       \hdherr |\bcov V|_{\backg}
    \Bigg\}
    \,.
    \label{13II18.2}
\end{eqnarray}
Using \eqref{12VIII1711}, we thus obtain
\ptcheck{14II18}
\begin{eqnarray}
 \nn
 \lefteqn{
  \mathcal{ A}^k \bcov_k V
  }
  &&
  \\
 &=&
 \gauge{
 \underbrace{
 -
  \left(
    \tensor{h}{^k_i} \bcheckpsi^i
    +
    \frac 12
    \phi\bcheckpsi^k
   \right)
   \bcov_k V
   }_{=:{\mathcal G}}
   }
   +
 \divergence{ \frac{1}{2}
      \bcov_\ell
       \left[
        \left(
         h^{k \ell} \phi
         +
         \frac{1}{4}
         \backg^{k \ell} \phi^2
        \right)
        \bcov_k V
       \right]
        \rsqrtbg
        }
\nn
  \\
   &&
       \quadratic{
       +
       \frac{1}{2}
        \left[
          \left(
           \frac{n}{4}
           +
           1
          \right)
          \lambda \phi^2
          -
          \frac{1}{4}
           \adsR{} \phi^2
           -
           h^{k \ell} \adsR{_{k \ell}} \phi
        \right]
         V}
         \nn
 \\
 &&
       \quadratic{
          +
          \hdherr
          |\bcov V|_{\backg}
         \rsqrtbg
        +
        \herr  {\absV}  \rsqrtbg
         }
       \,.
        \label{26IV18.9}
\end{eqnarray}
In the space-form case and using \eqref{20XI17.21},  \eqref{26IV18.9}   reads
 \ptcheck{14II18}
\begin{eqnarray}
 \nn
 \lefteqn{
  \mathcal{ A}^k \bcov_k V
 }
 &&
 \\
 &=&
 \gauge{
 \underbrace{
 -
  \left(
    \tensor{h}{^k_i} \bcheckpsi^i
    +
    \frac 12
    \phi\bcheckpsi^k
   \right)
   \bcov_k V
   }_{={\mathcal G}}
   } +
 \divergence{ \frac{1}{2}
      \bcov_\ell
       \left[
        \left(
         h^{k \ell} \phi
         +
         \frac{1}{4}
         \backg^{k \ell} \phi^2
        \right)
        \bcov_k V
       \right]
        \rsqrtbg
        }
        \nn
\\
 &&
       -
       \quadratic{
       \frac{1}{2}
        \left[
          \left(
           \frac{n}{4}
           +
           1
          \right)
          \phi^2 V
          +
          \hdherr
          |\bcov V|_{\backg}
        \right]
         \rsqrtbg
        +
        \herr  {\absV}  \rsqrtbg
         }
       .
        \phantom{xxxxx}
        \label{14II18.11}
\end{eqnarray}

\item
We can add the second term of \eqref{15VII17.11} to the first term of \eqref{14VII17.31}, namely $V\bR_{ij} g^{ij} \rsqrtbg $, using \eqref{12VIII1711}.
%
%
Thus, we have
%
%
\begin{eqnarray}
\lefteqn{
 V \bR_{ij} g^{ij} 
 +
  h_{ j \ell}
     (
 g^{ i j}  g^{ k \ell}  -
       g^{\ell j}  g^{ k i}
       )
     \bcov{_ i}\bcov{_ k}V \rsqrtbg
    }
 &&
 \nn
 \\
 &=&
 V \bR{}
 -
 h^{ij}
  \left(
    \bcov_i \bcov_j V + \lambda \backg_{ij} V
  \right)
 +
 \bR{_{ij}} \chi^{ij}
 \nn
 \\
 &&
 +
  h_{ j \ell}
  \left(
    \backg^{i j} \backg^{k \ell}
    -
    \backg^{ \ell j} \backg^{ k i}
    +
    \check{\chi}^{ijkl}
  \right)
  \bcov_i \bcov_k V
  \nn
  \\
&=&
  V
  \left[
   \bR{}
   +
   \left(
    \chi^{ij} + \hat{\chi}^{ij}
   \right)
   \bR{_{ij}}
   -
   \lambda
   \hat{\chi}^{ij} \backg_{ij}
  \right]\rsqrtbg
\nn
 \\
  &=&
  V
   \Big[
     \adsR{}
     -
     \tensor{h}{^i _\ell} h^{\ell j} \adsR{_{i j}}
     +
     h^{i j} \adsR{_{i j}} \phi
     +
     \adsR{} \sqrdof{h}
   \nn
   \\
   &&
     -
     \lambda
      \left(
        \phi^2
        +
        (n-2) \sqrdof{h}
      \right)
      +
      \herr
   \Big]
  \,,
  \label{19XII17.3}
\end{eqnarray}
where
\begin{eqnarray}
 \check{\chi}^{ i \ell j k}
  & := &
 2 \Big(
 -
 \backg^{ j [i}h^{\ell] k}
 -
 \backg^{k [\ell} h^{i] j}
 +
 h^{j [i} h^{\ell] k}
 +
 \backg^{j [i} \chi^{\ell] k}
\nn
\\
&&
 +
 \backg^{k [\ell} \chi^{i] j}
 -
 h^{j [i } \chi^{\ell ] k}
 -
 h^{k [\ell } \chi^{i ] j}
 +
 \chi^{j [i} \chi^{\ell ] k}
 \Big)
  \nn
\\
  & = &
 2 \left(
 -
 \backg^{ j [i}h^{\ell] k}
 -
 \backg^{k [\ell} h^{i] j}
 \right)
  + O(|h|^2)
 \,,
\end{eqnarray}
which possesses the algebraic symmetries of the Riemann tensor,
\begin{equation}
 \check{\chi}^{i \ell j k }
 =
 - \check{\chi}^{ \ell i j k }
 =
- \check{\chi}^{i \ell k j}
 =
 \check{\chi}^{j k i \ell }
 \,,
\end{equation}
and $\hat{\chi}^{i k} := h_{j\ell} \check{\chi}^{i \ell j k }$. Then, we have
\ptcheck{31I18}
\begin{eqnarray}
 \hat{\chi}^{i k} \backg_{i k}
 &=&
 h_{j \ell}
 \left(
  \backg^{j \ell} h^{i k}
  -
  \backg^{j i} h^{\ell k}
  -
  \backg^{k \ell} h^{i j}
  +
  \backg^{k i} h^{\ell j}
  +
  O\left(|h|^2_{\backg}  \right)
 \right)
 \backg_{ik}
 \nn
 \\
 &=&
 \phi^2 +( n- 2) |h|^2_{\backg} + O\left(|h |^3_{\backg} \right)
 \,.
 \label{18II18.3}
\end{eqnarray}
If $\backg$ is space-form, using \eqref{20XI17.21} and keeping in mind that $\lambda=-n$, \eqref{19XII17.3} becomes\ptcheck{31I18}
\begin{eqnarray}
 \lefteqn{
  V
  \left[
   \bR{}
   +
   \left(
    \chi^{ij} + \hat{\chi}^{ij}
   \right)
   \bR{_{ij}}
   -
   \lambda
   \hat{\chi}^{ij} \backg_{ij}
  \right]
  \rsqrtbg
  }
  &&
   \nn
\\
 & = &
  V
  \left[
   \bR
   +
   \chi^{ij} \bR_{ij}
   -
   (n-1+\lambda) \hat{\chi}^{ij} \backg_{ij}
   \right]
   \rsqrtbg
   \nn
   \\
   &=&
   V
    \bR
    \rsqrtbg
    +
    \quadratic{
    V
     \left[
    \phi^2
    -
     |h|^2_{\backg}
    +
    O\left(|h|_{\backg}^3 \right)
   \right]
   \rsqrtbg
   }
   \,.
   \label{25I18.2}
\end{eqnarray}

\end{enumerate}

Summarizing  we obtain, quite generally,
\begin{eqnarray}
 V \left( R - \bR \right) \rsqrtbg
 &=&
 \divergence{\overline{\mathcal{D}}}
  +
  \quadratic{\wmcQ }
    + \gauge{\mathcal{G}}\,,
  \label{1II18.1}
\end{eqnarray}
where
%
\begin{eqnarray}
 \overline{\mathcal{D}}
 &:=&
  \bcov{_ i}
   \big[
   V  g^{ m j}  g^{ i\ell }
    \left(
      \bcov{_ m} h_{ j\ell }
     -
     \bcov{_\ell } h_{ j m}
    \right)
    +
       (g^{ m j}  g^{ k i}
      - g^{ i  j}  g^{ km}
      )
     h_{ j m} \bcov{_ k}V
      \big]
\nn
 \\
  &&
  +
  \frac{1}{2}
  \bcov_i   \left[
   V
    \backg_{k \ell}
      \left(
        g^{j k} \bcov_j g^{i \ell}
        -
        g^{i k} \bcov_j g^{j \ell}
      \right)
   \right]
  \nn
  \\
  &&
   +
   \frac{1}{2}
     \bcov_i
  \left[
   \left(
  -3
   h^{i \ell} \tensor{h}{_\ell ^k}
  +
   g^{i k}
     |h|^2_{\backg}
   \right)
   \bcov_k V
  \right]
  \nn
\\
 && + \frac{1}{2}
      \bcov_i
       \left[
        \left(
         h^{k i} \phi
         +
         \frac{1}{4}
         \backg^{k i} \phi^2
        \right)
        \bcov_k V
       \right]
 \label{16II18.11}
   \rsqrtbg
\end{eqnarray}
is the sum of all divergence terms,
and where $\mathcal G$ is the gauge-dependent term defined in \eqref{26IV18.9}, which has no obvious sign but which can be made to vanish by a gauge transformation.
Finally,  $\wmcQ $ is the sum of quadratic terms and error terms, in the general case given by
\ptcheck{26II18}
\quadratic{
\begin{eqnarray}
 \wmcQ
 &:=&
 \Bigg\{
  -
  \frac{1}{4} g^{ i j} g^{ k p} g^{\ell q}
  \left(
     \bcov{_\ell } h_{ k p}   \bcov{_ q} h_{ i j}
     +
     \bcov{_ i} h_{ p q}   \bcov{_ j} h_{ k\ell }
  \right)
\nn
\\
&&
  +
  \frac{1}{2}
  \left(
    |\psi|^2_{\backg}
    -
    \chi^{i j} \adsR{_{i j}}
    +
    \tensor{h}{^i _\ell} h^{j m} \adsR{^\ell_{m i j}}
  \right)
\nn
  \\
   &&
   +
   \frac{3}{2}
     h^{i \ell} \tensor{h}{_\ell ^k} \adsR{_{i k}}
     -
      \frac{1}{2}
     \left[
       \adsR{}
       +
       \lambda \left( 3-n \right)
     \right]
     \sqrdof{h}
 \nn
 \\
 &&
      +
       \frac{1}{2}
        \left[
          \left(
           \frac{n}{4}
           +
           1
          \right)
          \lambda \phi^2
          -
          \frac{1}{4}
           \adsR{} \phi^2
           -
           h^{k \ell} \adsR{_{k \ell}} \phi
        \right]
\nn
  \\
   &&
     -
     \tensor{h}{^i _\ell} h^{\ell j} \adsR{_{i j}}
     +
     h^{i j} \adsR{_{i j}} \phi
     +
     \adsR{} \sqrdof{h}
 \nn
 \\
 &&
     -
     \lambda
      \left[
        \phi^2
        +
        (n-2) \sqrdof{h}
      \right]
   +
   \herr
   +
   \hdhsqerr \rsqrtbg
 \Bigg\}
   V
\nn
  \\
   &&
   +
     \hdherr
  | \bcov  V|_{\backg} \rsqrtbg
\nn
 \\
  &=&
  \Big\{
   -
   \frac{1}{4}
   \sqrdof{\bcov\phi}
   -
   \frac{1}{4}
   \sqrdof{\bcov h}
   +
   \frac{1}{2}
   \sqrdof{\psi}
   +
   \frac{1}{2}
   h^{i \ell} h^{j m} \adsR{_{\ell m i j}}
\nn
\\
&&
   +
   \frac{1}{2}
   h^{i j} \adsR{_{i j}} \phi
   +
    \frac{1}{2}
   \left[
     \adsR{}
     -
     \lambda (n-1)
   \right]
   \sqrdof{h}
\nn
  \\
   &&
    +
    \frac{1}{8}
    \left[
      \left(
       n -4
      \right)
      \lambda
      -
      \adsR{}
    \right]
    \phi^2
    +
    \herr
    +
    \hdhsqerr
 \Big\}V
\nn
\\
&&
 +
 \hdherr
  | \bcov  V|_{\backg}
  \,.
  \phantom{xxxxx}
   \label{26IV18.5}
\end{eqnarray}
}
For space-forms this becomes
{
\begin{eqnarray}
\nn
  \wmcQ
  &=&
  \Big[
  -
  \frac{1}{4} g^{ i j} g^{ k p} g^{\ell q}
  \left(
     \bcov{_\ell } h_{ k p}   \bcov{_ q} h_{ i j}
     +
     \bcov{_ i} h_{ p q}   \bcov{_ j} h_{ k\ell }
  \right)
  \nn
\\
 &&
  +
  \frac{1}{2}
  \left(
   |\psi|^2_{\backg}
   -
   \phi^2
   +
   n
   |h|^2_g
   +
   (3-n)
  |h|^2_{\backg}
   \right)
   +
 \phi^2
    -
     |h|^2_{\backg}
\nn
  \\
  &&  -
       \frac{1}{2}
          \left(
           \frac{n}{4}
           +
           1
          \right)
          \phi^2
    +
    \herr
    +
    \hdhsqerr
   \Big]
   V \rsqrtbg
\nn
  \\
   &=&
   \Big[
   -
    \frac{1}{4}
     |\bcov\phi|^2_{\backg}
     -
     \frac{1}{4}
     |\bcov h|^2_{\backg}
     +
     \frac{1}{2}
       |\psi|^2_{\backg}
       -
     \frac{n}{8}
      \phi^2
      +
      \frac 12
      |h|^2_{\backg}
  \nn
  \\
  &&
     +
     \herr
     +
     \hdhsqerr
    \Big]
    V \rsqrtbg
    \,.
     \phantom{xxxxx}
     \label{14II18.12}
\end{eqnarray}
}
Using   \eqref{20II18.1} and
\begin{equation}
 |h|^2_{\backg}
 =
 |\hat{h}|^2_{\backg}
 +
 \frac{1}{n} \overline{\phi}^2
 \,,
  \qquad
%
%
%
 |\bcov h|^2_{\backg}
 =
 |\bcov \hat{h}|^2_{\backg}
 +
 \frac{1}{n} |\bcov \,\overline{\phi}|^2_{\backg}
 \,,
\end{equation}
we can
rewrite $\wmcQ $ of \eqref{26IV18.5} in terms of the trace-free part of $h$ and of $\bcheckpsi$:
\begin{eqnarray}
\quadratic{
 \wmcQ
 }
 &=&
 \Big(%
  -
   \frac{n+2}{8n}
   |\bcov \, \phi|^2_{\backg}
   -
   \frac{1}{4}
   |\bcov \hat{h}|^2_{\backg}
   +
     \frac{1}{2}
    \hat h^{i \ell} \hat h^{j m}  \adsR{_{\ell m i j}}
    +
    \frac{n+2}{2n} \phi \hat h^{i j} \adsR{_{ i j}}
  \nn
  \\
    &&
      +
         \frac{n^2-4}{8n^2} \lambda
      \phi^2
    \gauge{
    +
    \frac{1}{2}
    \big(
       |\bcheckpsi|^2_{\backg}
       -
       \bcheckpsi^i\bcov_i \phi )
     }
     +
    \herr
    +
    \hdhsqerr
 \Big)
 V
 \nn
\\
 &&
 +
 \hdherr
  | \bcov  V|_{\backg}
  \,,
   \label{28II18.6}
\end{eqnarray}
where we used \eqref{12VIII1711}. Putting this into \eqref{1II18.1} we obtain \eqref{1II18.1a}. Also, when $\backg$ is a space-form metric, this gives
\begin{eqnarray}
 \mathcal{Q}
 &=&
  \Big[-
   \frac{n+2}{8n}
   |\bcov \, \phi|^2_{\backg}
   -
   \frac{1}{4}
   |\bcov \hat{h}|^2_{\backg}
   +
     \frac{1}{2}
    |\hat{h}|^2_{\backg}
    -
    \frac{ n^2-4}{8n}
    \phi^2
\nn
\\
&&
    +
    \herr
    +
    O(|h|_{\backg} |\bcov h|^2_{\backg})
  \Big]
     V \rsqrtbg
\nn
  \\
    &&
    \gauge{
    +
    \frac{1}{2}
    \big(
       |\bcheckpsi|^2_{\backg}
       -
       \bcheckpsi^i\bcov_i \phi )
       V
       \rsqrtbg
       }
          +
    \hdherr  |\bcov V|_{\backg} \rsqrtbg
    \,,
     \label{26IV18.3}
\end{eqnarray}
which is precisely \eqref{Sum26IV18.3}.

\appendix

\section{A weighted Poincar\'e inequality}
 \label{s26IV18.2}

When $\bcheckpsi =0$
all terms in \eqref{Sum26IV18.3} have the desired negative sign except for those involving undifferentiated occurrences of  $\hat h$. To address this, some integral identities will be needed.
Set
\begin{eqnarray}
 &
 ( {\overline{{\fancyD}}}\hat h)_{ijk}:= \frac 1 {\sqrt 2} \left(
  \bcov{}_i \hat h_{jk}
      -
  \bcov{}_j \hat h_{ik}
   \right)
   \,,
    \quad
  ({\overline{{\fancyL}}} v
  )_{ij}:= \frac 1 {2} \left(
  \bcov{}_i v_{j}
      +
  \bcov{}_j  v_{i}
   \right)
   \,,
   &
 \nonumber
\\
 \label{14II18.4}
  &
   (\overline{\mathrm{div}} \,\hat h)
   _j := -\bcov{}_i \hat h^{i}{}_{j}
    \,,
    \quad
     (\hat h_{dV})_i:= V^{-1}\hat h_{ij}\bcov^j V
     \,,
     &
\end{eqnarray}
and note that $\overline{{\fancyL}}^*=\overline{\mathrm{div}}$.
For any symmetric tensor $\check h$ we have (cf., e.g., \cite[Section~3]{DelayInversion})
\begin{equation}\label{14II18.3}
  (\overline{{\fancyD}}^*\overline{{\fancyD}}+\overline{{\fancyL}}\,{}\overline{{\fancyL}}^*)
  \check h
  =(\bcov^*\bcov+\overline{\Ric}-\overline{\mathrm{Riem}})
  \check h
  \,,
\end{equation}
where
\begin{equation}\label{15II18.1}
  [(\overline{\Ric} - \overline{\mathrm{Riem}}){\hat h}]_{ij}=\frac{1}{2}(\overline{R}_{ik}{\hat h}^k{}_j+\overline{R}_{jk}{\hat h}^k{}_i
   -2\overline{R}_{ikj\ell}{\hat h}^{k\ell})
  \,.
\end{equation}

Assume, first, that $\backg$ is a space-form.
Multiplying \eqref{14II18.3} by $V{\hat h}$ and integrating by parts, after some simple manipulations one obtains
\begin{eqnarray}
   \int  |  {\hat h}|_{\backg{}}^2 V \sqrtbg
   & = &
    \frac 1 {n+1}
    \int \Big[
     (|\bcov {\hat h}|_{\backg{}}^2-|\overline{{\fancyD}} {\hat h}|_{\backg{}}^2 - |\overline{\mathrm{div}}\, {\hat h}|_{\backg{}}^2
    )V
    \nn
    \\
    &&
     \qquad +
     \bcov_j({\hat h}_{ik}\bcov^iV {\hat h}^{jk}) - 2 {\hat h}^{ik} \bcov_iV \bcov^j {\hat h}_{jk}
     \Big]
     \sqrtbg
     \nn
\\
   & = &
    \frac 1 {n+1}
    \int \Big[
     (|\bcov {\hat h}|_{\backg{}}^2-|\overline{{\fancyD}} {\hat h}|_{\backg{}}^2
      - |\overline{\mathrm{div}}\, {\hat h} - {\hat h}_{dV}|_{\backg{}}^2
     )V
     \nn
     \\
     &&
     \qquad
     +
     \bcov_j({\hat h}_{ik}\bcov^iV {\hat h}^{jk}) + |{\hat h}_{dV}|^2_{\backg} V
     \Big]
     \sqrtbg
  \,.\label{14II18.3+}
\end{eqnarray}

In a coordinate system in which the (suitably-normalised) anti-de Sitter metric $\overline{\mathbf g}$ reads
\begin{equation}\label{16II18.1}
  \overline{\mathbf{g}} = - (r^2 +1) dt^2 + \frac{dr^2}{r^2+1} + r^2 d\Omega^2
\end{equation}
we choose $V$ as in \eqref{17XII18.1} with $|\vec A|  \le A^0$
so that
\begin{equation}\label{16II18.3}
  |dV|_{\backg} < V
    \quad
  \Longrightarrow
  \quad
  |{\hat h} _{dV}|_{\backg} \le |{\hat h} |_{\backg}
  \,.
\end{equation}
This gives
\begin{eqnarray}
   \int  |  {\hat h}|_{\backg{}}^2 V
   \sqrtbg
   & \le  &
    \frac 1 {n}
    \int \Big[
     (|\bcov {\hat h}|_{\backg{}}^2-|\overline{{\fancyD}} {\hat h}|_{\backg{}}^2 - |\overline{\mathrm{div}}\, {\hat h} - {\hat h}_{dV}|_{\backg{}}^2
     )V
     \nn
     \\
     &&
     \qquad +
     \bcov_j({\hat h}_{ik}\bcov^iV {\hat h}^{jk})
     \Big]
     \sqrtbg
  \,.\label{17XII18.9}
\end{eqnarray}
which provides the desired \emph{weighted Poincar\'e inequality} for space-forms when the trace-free tensor field $\hat h$ decays sufficiently fast so that the divergence term gives no contribution:
\begin{eqnarray}
   \int  |  {\hat h}|_{\backg{}}^2 V
   \sqrtbg
   & \le  &
    \frac 1 {n}
    \int
     |\bcov {\hat h}|_{\backg{}}^2
     \sqrtbg
  \,.\label{14II18.3++}
\end{eqnarray}

We now indicate how to adapt the above argument to the general case, without assuming that the metric is a space form. For this, multiplying \eqref{14II18.3} by $V {\hat h}$ and integrating by parts  we obtain
\begin{eqnarray}
 \lefteqn{
   \int
    \left(
      \adsR{_{i k j \ell}} {\hat h}^{k \ell}
      -
      \adsR{_{i k}} \tensor{{\hat h}}{^k_j}
    \right)
    {\hat h}^{i j}      V \sqrtbg
    }
    &&
    \nn
    \\
    &&
   =
    \int \Big[
     (|\bcov {\hat h}|_{\backg{}}^2-|\overline{\fancyD} {\hat h}|_{\backg{}}^2 - |\overline{\mathrm{div}}\, {\hat h}|_{\backg{}}^2
    )V
    +
     \bcov_j({\hat h}_{ik}\bcov^iV {\hat h}^{jk})
\nn
  \\
    &&
    \qquad
    - 2 {\hat h}^{ik} \bcov_iV \bcov^j {\hat h}_{jk}
     -
     \left(
       \adsR{_{i j}} {\hat h}^{j k} \tensor{{\hat h}}{^i_k}
       -
       \lambda  \sqrdof{{\hat h}}
     \right)
     V
     \Big]
     \sqrtbg
     \,.
      \label{3III18.2}
\end{eqnarray}
To continue, it is convenient to introduce
\begin{equation}\label{3III18.1}
  \bhatpsi_i := -\bcov_j  h^j{}_i
  + \frac 12 \bcov_i \overline \phi
   = -\bcov_j {\hat h}^j{}_i
  + \frac {n+2}{2n} \bcov_i \overline \phi
\end{equation}
(note that this differs from $\bcheckpsi_i$ by higher order terms).
In this notation \eqref{3III18.2} can be rewritten as
\begin{eqnarray}
 \nn
  \lefteqn{
   \int
      \left(
       \adsR{_{i k j \ell}} {\hat h}^{k \ell} {\hat h}^{i j}
       -
       \lambda  \sqrdof{{\hat h}} \right)
       V \sqrtbg
       }
       &&
\\
 &   =
  &
    \int \Big[
      \Big( |\bcov {\hat h}|_{\backg{}}^2-|\overline{\fancyD} {\hat h}|_{\backg{}}^2
       - |\bhatpsi -  \frac {n+2}{2n} \bcov\, \overline \phi|_{\backg{}}^2
 \nn
\\
 &&
       +2 (\bhatpsi^k  - \frac {n+2}{2n} \bcov^k \overline \phi)({\hat h}_{d V})_k
    \Big) V
    +   \bcov_j({\hat h}_{ik}\bcov^iV {\hat h}^{jk})
     \Big]
     \sqrtbg
     \,.\label{28II18.30}
\end{eqnarray}
One should keep in mind that the divergence term at the right-hand side is irrelevant for many purposes, in that it gives a vanishing contribution for suitably decaying fields when the integral in \eqref{28II18.30} is taken over the whole manifold.

Let $\gamma>0$ be a constant, which might have to be chosen on a case-by-case basis depending upon the background geometry at hand. The trivial identity
\begin{equation}\label{28II18.31}
  -  2\bcov^k {} \overline \phi ({\hat h}_{d V})_k
   =  - | \gamma^{-1} \bcov \,
    \overline \phi + \gamma {\hat h}_{d V}  |^2_{\backg}
  +   | \gamma^{-1} \bcov  {} \, \overline \phi |^2_{\backg}  + |\gamma {\hat h}_{d V}  |^2_{\backg}
\end{equation}
leads to the following version of
\eq{28II18.30}:
\begin{eqnarray}
 \nn
  \lefteqn{
   \int
      \left(
       \adsR{_{i k j \ell}} {\hat h}^{k \ell} {\hat h}^{i j}
       -
       \lambda  \sqrdof{{\hat h}}
       {- \frac {n+2}{2n} } |\gamma {\hat h}_{d V}  |^2_{\backg} \right)
       V \sqrtbg
       }
       &&
\\
 &   =
    \displaystyle \int \Big[
  &  \Big( |\bcov {\hat h}|_{\backg{}}^2-|\overline{\fancyD} {\hat h}|_{\backg{}}^2
  - |\bhatpsi^k  - \frac {n+2}{2n} \bcov  \, \overline\phi|_{\backg{}}^2
       +2  \bhatpsi^k  ({\hat h}_{d V})_k
        \nn
       \\
       &&
        {- \frac {n+2}{2n} } | \gamma^{-1} \bcov  \phi + \gamma {\hat h}_{d V}  |^2_{\backg}
   {+  \frac {n+2}{2 n}} | \gamma^{-1} \bcov  \, \overline\phi |^2_{\backg}
    \Big) V
    \nn
\\
 &&
    +   \bcov_j({\hat h}_{ik}\bcov^iV {\hat h}^{jk})
     \Big]
     \sqrtbg
     \,.
     \label{28II18.25}
\end{eqnarray}
\emph{Suppose} that
there exist  constants $c\ge 0$ and $\varepsilon>0$ such that    for all $\overline \phi$ and
${\hat h}  $ we have
\begin{multline}
\label{28II18.33}
    \frac{1}{2}
    {\hat h}^{i \ell} {\hat h}^{j m}  \adsR{_{\ell m i j}}
    +
    \frac{n+2}{2n} \overline \phi {\hat h}^{i j} \adsR{_{ i j}}
      +
         \frac{n^2-4}{8n^2} \lambda
      \overline  \phi^2
     \\
     \le c
     \left(
       \adsR{_{i k j \ell}} {\hat h}^{k \ell} {\hat h}^{i j}
       -
       \lambda  \sqrdof{{\hat h}}
       -  \frac {n+2}{2n}
       |\gamma {\hat h}_{d V}  |^2_{\backg}
       \right)
       {
       - \varepsilon \sqrdof{{\hat h}}
       }
       \,.
\end{multline}
Integrating \eqref{Sum28II18.6}  over the manifold in the gauge $
\bcheckpsi \equiv 0$,
using  \eqref{28II18.25}-\eqref{28II18.33}  and the decay conditions on $h$  we obtain
%
\begin{eqnarray}
\int
 \wmcQ \sqrtbg
 &\le
 &
 \int
 \Big \{
 \Big[
  -
   \frac{n+2}{8n}
   |\bcov \, \phi|^2_{\backg}
   -
   \frac{1}{4}
   |\bcov \hat{h}|^2_{\backg}
       - \varepsilon \sqrdof{{\hat h}}
   \nn
   \\
   &&  +
    \herr
    +
    \hdhsqerr
 \Big]V
 +
 \hdherr
  | \bcov  V|_{\backg}
  \nn
\\
&&
+ c
      \Big( |\bcov {\hat h}|_{\backg{}}^2
  + \frac {n+2}{2n} \big( \gamma^{-2} - \frac {n+2}{2n} \big) | \bcov  \phi |^2_{\backg}
    \Big) V
  \Big\}
     \sqrtbg
     \,.
   \label{28II18.6+}
\end{eqnarray}
The right-hand side will be strictly negative, as desired, for all sufficiently small  $\|h\|_{L^\infty}$ and $\|\bcov   h\|_{L^\infty}$, provided that  $V>0$, that $V^{-1}|\bcov  V|_{\backg}$ is bounded, and that
\begin{equation}\label{28II18.29}
0<  c< \frac 14
  \,,
  \quad
  c   \big(\gamma^{-2} -  \frac {n+2}{2n}\big) <
  \frac 14 %
   \,.
\end{equation}
This reduces the positivity issue to the algebraic inequality  \eqref{28II18.33},  with $\gamma$ and $c$ satisfying \eqref{28II18.29}. The existence of $c$, and its value, has to be checked on a case-by-case basis.
We note that this strategy does not allow one to conclude in the case of Horowitz-Myers instantons.

\medskip

\noindent{\sc Acknowledgements:} PTC wishes to thank IHES, Bures-sur-Yvette, for hospitality and financial support during part of work on this paper. Useful comments from Erwann Delay are acknowledged.

\providecommand{\bysame}{\leavevmode\hbox to3em{\hrulefill}\thinspace}
\providecommand{\MR}{\relax\ifhmode\unskip\space\fi MR }
\providecommand{\MRhref}[2]{%
  \href{http://www.ams.org/mathscinet-getitem?mr=#1}{#2}
}
\providecommand{\href}[2]{#2}

\end{document}